\documentclass[dvipsnames]{aa} 
\usepackage{graphicx}
\graphicspath{{figures}}
\usepackage{subfigure}
\usepackage{txfonts}
\usepackage{amssymb}
\usepackage{amsmath}
\usepackage{natbib}
\usepackage{xcolor}
\usepackage{url}
\usepackage{placeins}
\usepackage{multirow} 
\usepackage{hyperref}
\hypersetup{
  colorlinks=true,
  allcolors=blue
}
\DeclareMathAlphabet{\mathitbf}{OML}{cmm}{b}{it}

\newcommand{\By}{\mathit{B_y}}
\newcommand{\Bz}{\mathit{B_z}}

\newcommand{\Ef}{E_{\rm F}}

\newcommand{\Hj}{H_{\rm J}}

\newcommand{\mxmx}{{\rm Mx}$^2$}
\newcommand{\mx}{{\rm Mx}}
\newcommand{\gauss}{{\rm G}}
\newcommand{\erg}{{\rm erg}}

\newcommand{\kmps}{{$\mathrm{~km\,s}^{-1}$}}

\newcommand{\kmsps}{{$\mathrm{~km}^2~\mathrm{s}^{-1}$}}
\newcommand{\mxps}{{$\mathrm{~Mx}~\mathrm{s}^{-1}$}}

\newcommand{\keV}{{$\mathrm{~keV}$}}

\begin{document}

\title{Tracking magnetic flux and helicity from Sun to Earth}
  \subtitle{Multi-spacecraft analysis of a magnetic cloud and its solar source}
\author{J.~K. Thalmann\inst{1}
    \and M. Dumbovi\'c\inst{2}
    \and K. Dissauer\inst{4}
    \and T. Podladchikova\inst{3}
    \and G. Chikunova\inst{3}
    \and \\
    M. Temmer\inst{1}
    \and E. Dickson \inst{1}
    \and A.~M. Veronig\inst{1,3}
}

\institute{University of Graz, Institute of Physics, Universit\"atsplatz 5, 8010 Graz, Austria\\   
  \email{julia.thalmann@uni-graz.at}\\
  \and Hvar Observatory, Faculty of Geodesy, University of Zagreb, Kaciceva 26, HR-10000 Zagreb, Croatia\\
  \and Skolkovo Institute of Science and Technology, Bolshoy Boulevard 30, bld. 1, Moscow 121205, Russia\\
  \and NorthWest Research Associates, 3380 Mitchell Lane, Boulder, CO 80301, USA\\
  \and University of Graz, Kanzelh\"ohe Observatory for Solar and Environmental Research, Kanzelh\"ohe 19, 9521 Treffen, Austria\\
}
    
\date{Received June 13, 2022 ; accepted September 30, 2022}
\abstract
{} 
{
We analyze the complete chain of effects --- from Sun to Earth --- caused by a solar eruptive event in order to better understand the dynamic evolution of magnetic-field related quantities in interplanetary space, in particular that of magnetic flux and helicity.
}
{
We study a series of connected events --- a confined C4.5 flare, a flare-less filament eruption and a double-peak M-class flare --- that originated in NOAA active region (AR) 12891 on late 2021 November 1 and early 2021 November 2. We deduce the magnetic structure of AR 12891 using stereoscopy and nonlinear force-free (NLFF) magnetic field modeling, allowing us to identify a coronal flux rope and to estimate its axial flux and helicity. Additionally, we compute reconnection fluxes based on flare ribbon and coronal dimming signatures from remote sensing imagery. Comparison to corresponding quantities of the associated magnetic cloud (MC), deduced from {\it in-situ} measurements from {\it Solar Orbiter} and near-Earth spacecraft, allows us to draw conclusions on the evolution of the associated interplanetary coronal mass ejection. The latter are aided through the application of geometric fitting techniques (graduated cylindrical shell modeling; GCS) and interplanetary propagation models (drag based ensemble modeling; DBEM) to the interplanetary CME.
}
{NLFF modeling suggests the host AR's magnetic structure in the form of a left-handed (negative-helicity) sheared arcade/flux rope reaching to altitudes of 8–10 Mm above photospheric levels, in close agreement with the corresponding stereoscopic estimate. Revealed from GCS and DBEM modeling, the ejected flux rope propagated in a self-similar expanding manner through interplanetary space. Comparison of magnetic fluxes and helicities processed by magnetic reconnection in the solar source region and the respective budgets of the MC indicate a considerable contribution from the eruptive process, though the pre-eruptive budgets appear of relevance too.
}
{} 
\maketitle

\section{Introduction}
\label{intro}

Earth-directed solar coronal mass ejections (CMEs; magnetized clouds of coronal plasma, and sometimes also cooler and denser material originating from the lower solar atmosphere)  ejected from the solar atmosphere to interplanetary space, are known to severely impact our space weather \citep[e.g., reviews by][]{2007LRSP....4....1P,Koskinen2017,2021LRSP...18....4T}. The arrival of CME-driven shocks and their associated material are registered in {\it in-situ} observations by near-Earth spacecraft  in the form of sudden increases in solar wind speed and plasma-$\beta$, in combination with enhanced plasma density and temperature, as well as a drastically enhanced magnetic field strength at the shock boundary \citep[e.g., review by][]{2009SSRv..144..351V}. The interplanetary manifestations of CMEs are usually referred to as interplanetary CMEs (ICMEs). 
A particular subset of ICMEs, so-called magnetic clouds (MCs), adhere to an inherent magnetic field characteristic for a magnetic flux rope, i.e., being twisted around a common axis \citep[e.g.,][]{1981JGR....86.6673B}. Characteristic {\it in-situ} signatures of MCs include a smooth rotation of enhanced magnetic field, low proton temperatures and plasma-$\beta$ \citep{2006SSRv..123...31Z}. MC-like features are present in $\approx$\,77\% of all ICMEs while the remaining (``non-flux-rope'') events tend to exhibit more internal complexity \citep[e.g.,][]{2019SoPh..294...89N}.

When traveling through interplanetary space, ICMEs are expanding in a manner determined by the interaction of their inherent magnetic field and that of the ambient solar wind \citep[e.g.,][]{2009A&A...498..551D}. In general, ICMEs tend to expand self-similarly in the radial direction \citep[e.g.,][]{2019ApJ...877...77V}. Therefore, the increase in their size as well as the corresponding decrease in the magnetic field strength can be described by a power-law function \citep{bothmer98, leitner07,demoulin08,2010A&A...509A..39G,2019JGRA..124.4960G,2020JGRA..12527084S,2022ApJ...933..127D}. Observational  studies constrain the size and magnetic field power-law indices to 0.45\,$<$\,$n_a$\,$<$\,1.14 and $-1.89$\,$<$\,$n_B$\,$<$\,$-0.88$, respectively \citep[e.g.,][]{gulisano12}. More specifically, a study by \cite{2016A&A...595A.121P} has shown that the best-fit power law index to describe the decrease in the ICME magnetic field magnitude up to 1~AU is $n_B$\,$=$\,$-1.6$. Regardless of the expansion, the magnetic flux of the MC will be conserved under ideal MHD conditions, whereas it may not be conserved in the case of magnetic reconnection between the MC structure and the ambient solar wind \citep{manchester17}.

The rotational profile of the magnetic field of a MC indicates its handedness (geometrical sense) \citep[][]{bothmer98,2018SpWea..16..442P}, thus is indicative of its magnetic helicity. The comfortable property of magnetic helicity to be quasi-conserved even in the case of high Reynolds numbers \citep{1984GApFD..30...79B} implies that a consistency regarding the magnetic helicity budget (both, in sign and magnitude) is to be revealed when tracing knowingly associated features in the solar atmosphere and in interplanetary space. In other words, the helicity budget from a (pre-eruptive) solar source region has to roughly match that of an associated ICME/MC measured near Earth. In a pioneering study of 46 solar eruptions associated with sigmoidal structures observed in soft X-rays (SXRs), \cite{2002JGRA..107.1234L} found an overall positive correlation between the shape of coronal sigmoids and the handedness of the associated MCs. In the systematic study of helicity in MCs and their associated solar source regions for 12 events, \cite{2004JGRA..109.5106L} found the helicity of MCs to be typically an order of magnitude greater than that of the corresponding host active region (AR) (estimates based on linear force-free models of the coronal magnetic field) and no systematic sign or amplitude relationship between them. This led them to conclude that only magnetic reconnection in the eruptive process between the active-region and overlying magnetic field can explain the resulting MC helicity.

Also the magnetic flux of a MC can be set in relation to the magnetic flux involved in the magnetic reconnection process that caused the expulsion of the coronal plasma from a solar source region. During a large eruptive flare, the outward erupting CME leaves behind a growing magnetic field arcade (emitting in SXRs and at extreme ultra-violet (EUV) wavelengths) that is anchored at chromospheric locations of enhanced H$\alpha$ and UV emission, the latter separating from each other and the polarity inversion line (PIL) as time progresses. This enhanced emission is caused by particles accelerated towards the solar surface along newly reconnected field, upon deposition of their energy when interacting with the ambient chromospheric plasma \citep[for reviews see][]{2011SSRv..159...19F,Benz2017,Green2018}. As it is generally accepted that the flare-induced acceleration can only stem from magnetic reconnection, flare ribbons can be used to trace the local reconnection rate \citep[thus, the flare reconnection flux;][]{2002A&ARv..10..313P}. Similarly, coronal dimmings \citep[][]{1998GeoRL..25.2465T,2000GeoRL..27.1431T, 2007ApJ...659..758Q,dissauer+2018b_apj,dissauer+2019_apj} can be used for an estimate of the global reconnection rate as they reflect plasma evacuation in the low corona along of field lines of an arcade initially overlying a pre-existing flux rope and closed down by magnetic reconnection in the wake of a CME. In that case, coronal dimming develops ahead in time of magnetic reconnection. Furthermore, when magnetic reconnection happens at large coronal altitudes and is not energetic enough to produce visible radiation signatures in near-surface layers of the solar atmosphere, the magnetic flux encompassed by dimming areas was suggested to represent a better estimate of the reconnected flux than that encompassed by flare ribbons \citep[e.g.,][]{Forbes2000, 2004ApJ...602..422L, 2007ApJ...659..758Q}. 

\cite{2007ApJ...659..758Q} systematically summarized how the low-corona reconnection flux may be related to the magnetic flux of ICMEs/MCs. When magnetic reconnection takes place below a pre-existing magnetic flux rope as in the 2.5D standard flare model \citep[see, e.g.,][for details]{2004ApJ...602..422L}, it contributes solely to the poloidal (azimuthal) component of the ejected flux rope. Here, the poloidal flux refers to the integration of the magnetic field projected to a plane that is perpendicular to the flux-rope axis, hence related to the amount of twist along of the axis of the flux rope. Then, the poloidal flux of an associated MC should exceed the (dimming) reconnection flux. In the scenario of "{\it in-situ}-formed" flux ropes, where twisted magnetic flux ropes form from sheared arcades and then erupt, the entirety of its flux is anchored to the solar surface. Then, the poloidal flux should be close to the flare reconnection flux.

The study of \cite{2004JGRA..109.5106L} revealed a close correspondence between the axial magnetic fluxes of MCs (approximated using a linear force-free field solution in cylindrical geometry) and their host AR (estimated by spatial integration of the vertical magnetic field magnitude), with their ratio tending to be of order unity. \cite{2007ApJ...659..758Q} obtained that the poloidal/toroidal magnetic flux budget of MCs is comparable to/a fraction of the reconnection flux (measured based on areas in optical, UV, and EUV observations swept by flare ribbons). This is frequently interpreted as evidence of the formation of the helical structure of a magnetic flux rope by reconnection, in the course of which magnetic flux as well as helicity is added \citep[for individual case studies see, e.g.,][]{2006SoPh..238..117A,2007SoPh..244...45L,2008AnGeo..26.3139M,Moestl2009,2017SoPh..292...93T}. Depending on the relative magnitude of the magnetic flux and helicity estimates, different conclusions may be drawn regarding the ICME passage in interplanetary space. Reduced magnetic fluxes and helicities recovered from MC analysis might indicate an erosion while the ICME propagates through interplanetary space \citep[e.g.,][]{2006A&A...455..349D,2008AnGeo..26.3139M,2015JGRA..120...43R,2017SoPh..292...93T}. Corresponding interpretations, however, are to be considered with care due to the significant uncertainties in the underlying estimates.

Active region 12891 was the first geo-effective AR of the current solar cycle 25 when it produced a long-duration M-class flare (SOL-2021-11-02T01:20M1.7; peak time $\sim$03:01\,UT; heliographic position $\sim$N16W09), preceded by a smaller C-class flare (SOL-2021-11-01T23:35C4.5; peak time $\sim$23:40\,UT; heliographic position $\sim$N17E04). An extended filament partially erupted during successive activity and gave rise to a CME and associated MC arrival at Earth. The ICME was clearly captured by the {\it Solar and Heliospheric Observatory Large Angle and Spectrometric Coronagraph Experiment} and {\it Solar Terrestrial Relations Observatory Ahead} imagery, separated by $-34.3$~degrees from the Sun-Earth line. The MC caused clearly identifiable \textit{in-situ} signatures measured near Earth, by the almost aligned {\it Solar Orbiter} spacecraft (located 3~degrees East in longitude from Earth at a distance of $\sim$0.84~AU from the Sun) and the {\it Advanced Composition Explorer} and {\it Wind} satellites located at L1 (at a distance of $\sim$0.98~AU). This setting allows us to trace physical parameters from the Sun to {\it Solar Orbiter} and further to Earth-orbit, including magnetic fluxes and helicities, under the assumption that the ejected flux rope did not change its magnetic morphology during the transit through interplanetary space. Our aims in this paper are twofold. First, state-of-the-art modeling is applied to retrieve the structure of the (pre-)eruptive active-region coronal magnetic field by means of stereoscopy and nonlinear force-free magnetic field modeling. Second, the flare reconnection flux as well as magnetic helicity in the solar source region are estimated based on remote-sensing imagery and compared to corresponding estimates of the associated MC based on \textit{in-situ} measurements, in order to allow for interpretations of the Sun-Earth connection and ICME evolution.

\begin{figure*}
\centering
\includegraphics[width=0.9\textwidth]{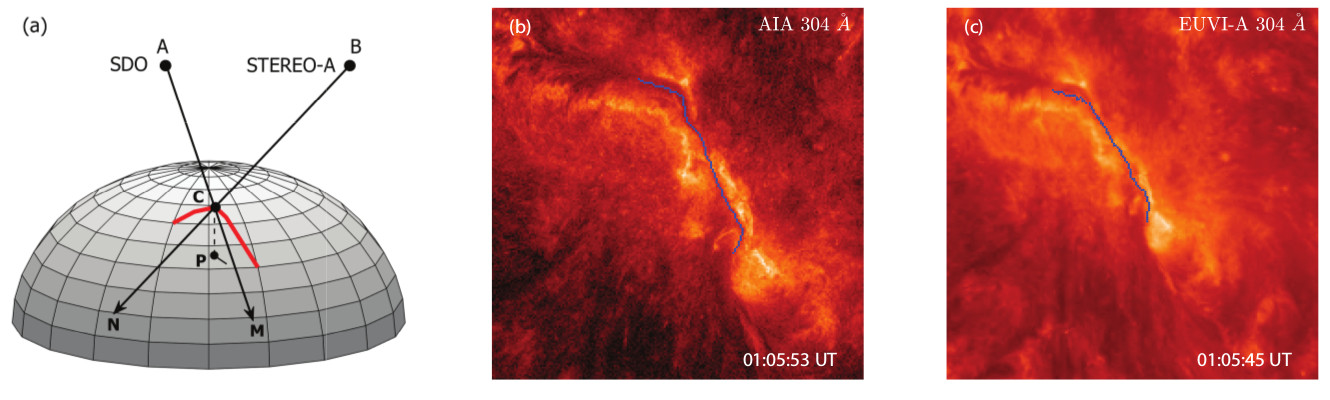}
\caption{3D reconstruction of the extended filament. (a) Schematic illustration of triangulation and a filament (red) above the solar surface. Points A and B show schematic locations of the SDO and STEREO-A spacecraft, respectively. C is the highest point of the filament. Point P shows the orthogonal projection of C onto the sphere. Points M and N are the projections of point C on the sphere along the LOS of SDO (AC) and STEREO-A (BC). (b) Selected points M (blue) in the AIA 304\,\AA~image along the filament axis are used for the 3D reconstructions. (c) STEREO-A image showing the points N (blue) matching the same features observed in the SDO image.
}
\label{fig:method_3d}
\end{figure*}

\section{Data and Methods}
\label{s:data}

\subsection{Coronal aspects} 

\subsubsection{Structural properties} 
\label{sss:methods_structure}

Multi-point observations from the Solar Dynamics Observatory \citep[SDO;][]{2012SoPh..275....3P} Atmospheric Imaging Assembly \citep[AIA;][]{2012SoPh..275...17L} and Solar Terrestrial Relations Observatory-Ahead \citep[STEREO-A;][]{Kaiser:2008} Extreme-Ultraviolet Imaging Telescope \citep[EUVI;][]{Wuelser2004} allow us to use stereoscopy for 3D reconstructions of the extended filament and to estimate its height using tie-pointing and triangulation techniques \citep{Inhester2006,Liewer2009}. Figure\,\href{fig:method_3d}{\ref{fig:method_3d}(a)} shows the triangulation scheme applied to a theoretical filament (indicated by the red curve) above the solar surface. Points A and B schematically indicate the locations of the SDO and STEREO-A spacecraft, respectively. The elevation of a certain point along of the filament (indicated by "C") is defined by the distance CP, where P is the orthogonal projection of C onto the solar sphere. C is observed by both spacecraft and is located at the intersection of the respective view directions. The line-of-sight (LOS) from SDO (at position "A") directed to point C intersects the solar surface in point "M". Similarly, the LOS of STEREO-A (at position "B") directed to point C intersects the solar surface in point "N". To construct the paths AM and BN in 3D and to define their point of intersection (C), identical features along the filament as observed from SDO (blue dots in Figure\,\href{fig:method_3d}{\ref{fig:method_3d}(b)}) and from STEREO-A (blue dots in Figure\,\href{fig:method_3d}{\ref{fig:method_3d}(c)}) are matched using epipolar geometry \citep[for further details we refer to][]{Podladchikova2019}. 

Being tightly related to the observed structure of the solar corona, the magnetic field in and around a solar AR can be indirectly inferred from extrapolation of magnetic field measurements at a photospheric level to the coronal volume. To do so, we use Cylindrical Equal Area (CEA) projected photospheric magnetic field vector data with the azimuthal component of the vector magnetic field being disambiguated \citep[][]{1994SoPh..155..235M,2009SoPh..260...83L} binned to a plate scale of 720~km as an input to a nonlinear force-free (NLFF) model \citep{2012SoPh..281...37W}. Having the modeled 3D magnetic field structure at hand, we are able to retrieve structural properties associated to the observed filament channel, such as the coronal altitude of the flux rope axis and arcade field which can be compared to the stereoscopy-based estimates. Also, magnetic-field related properties, such as the axial magnetic flux, average magnetic field, etc., can be estimated from the NLFF modeling, and compared to corresponding estimates for the {\it in-situ} measured/deduced MC properties (see Section\,\ref{ss:methods_insitu}).

In order to gain information on the time evolution of the non-potential coronal magnetic field in and around AR~12891, we construct several time series of NLFF models using time series of vector magnetograms at 12-minute time cadence around the time of the M-class flares and a 1-hour cadence otherwise. From those models and the corresponding potential (current-free) model fields, the free magnetic energy ($\Ef$) can be readily estimated. These NLFF model time series are force- and divergence-free (solenoidal) to a different level based on different combinations of free model parameters \citep[for a dedicated in-depth study see][]{2020A&A...643A.153T}. With the very level of solenoidality of the NLFF solutions being utterly important for the reliability of subsequent computation of magnetic helicity \citep{2013A&A...553A..38V,2016SSRv..201..147V,2019ApJ...880L...6T}, we select the NLFF time series of highest solenoidal quality (with non-solenoidal errors of less than 30\% of the free magnetic energy; for a dedicated study see \cite{2019ApJ...880L...6T}) for subsequent analysis and apply the Coulomb-gauge finite-volume helicity method of \cite{2011SoPh..272..243T} to compute a physically meaningful magnetic helicity of the coronal volume \citep[the "relative" helicity;][]{1984JFM...147..133B,1984CPPCF...9..111F}. In particular, we compute the relative helicity of the current-carrying field \citep[$\Hj$;][]{1999PPCF...41B.167B, 2003and..book..345B}, presumably being mostly determined by the magnetic structure hosting the observed filament. 

\subsubsection{Flare, CME and dimming analysis}
\label{sss:observations}

\subsubsection*{\it Thermal and non-thermal flare emission}

We analyze the thermal and non-thermal X-ray emissions using the X-ray sensor (XRS) of the Geostationary Operational Environmental Satellites (GOES) and the Spectrometer Telescope for Imaging X-rays \cite[STIX;][]{Krucker2020} onboard {\it Solar Orbiter} \citep[][]{muller20} to derive the characteristics of the flare-accelerated electron beams and to study the response of the ambient plasma to that energy input.

STIX level-1 compressed pixel science data and level-4 spectrogram science data were used to generate light curves, spectra and images (for a description of the STIX data compression levels, we refer to \cite{Krucker2020}). The STIX data is observed in 32 science energy channels from 4--150\,\keV, with a spectral resolution of 1\,\keV\ up to 16\,\keV. For the present study, 17 channels up to 28\,\keV\ showed significant signal above background and were used for analysis. The spectrogram data is binned over all pixels and detectors on board. Therefore, it has a much lower telemetry requirement and can be downloaded at the full observed time cadence which is as high as 0.5~seconds during a flare. The pixel data on the other hand is needed to produce images. As each pixel is downloaded separately it has less uncertainty due to statistics and compression but is only available at a reduced time cadence. 

\subsubsection*{\it Flare ribbons}

Flare ribbons are best observed in AIA filters capturing emission from the low solar atmosphere, like 1700\,\AA\ (photospheric), 1600\,\AA\  (upper photosphere and transition region), and 304\,\AA\ (chromosphere and transition region). They can be also observed in the coronal AIA filters, yet is their identification more difficult and ambiguous  since also flare loops and arcades appear brightened in these filters (partly also in the 304\,\AA\ channel). Therefore, we use sequences of AIA 1600\,\AA\ maps to identify the flare ribbons, corrected for differential rotation to a reference time of 01~November 23:05\,UT. 

To identify and segment the flare ribbon pixels, we apply a threshold-based method, following \cite{veronig2015} and \cite{2018ApJ...853...41T}. More precisely, we determine the lowest intensity maximum $I_m$ across the entire set of AIA maps, which usually corresponds to a time of low solar activity. Empirically, we find a scale factor $1.2\,I_m$ as a threshold level suitable to extract both, ribbons associated to the C4.5 as well as to the M-class flares. To minimize artifacts due to saturated pixels and blooming around the flare peak, a requirement for a flare pixel to be identified is to be detected as such in at least five consecutive images \cite[cf.\,][]{thalmann2015}. Using co-registered Helioseismic and Magnetic Imager \citep[HMI;][]{2012SoPh..275..229S} line-of-sight (LOS) magnetic field maps, we derive the total magnetic reconnection flux from the derived flare ribbon masks. To estimate the uncertainty of the reconnection flux due to the specific threshold used, we apply a $\pm$5\% change to the threshold level, and compute the mean values and corresponding standard deviation of the reconnection flux for further analysis.

\subsubsection*{\it Coronal Dimmings}

For the analysis of coronal dimmings, we use sequences of AIA 211\,\AA\ maps corrected for differential rotation with respect to the same reference time as for the flare ribbon analysis (01~November 23:05\,UT). This choice is based on the systematic study performed by \cite{dissauer+2018b_apj} who revealed that 211\,\AA\  and 193\,\AA\  are optimal for the observation and extraction of coronal dimmings, as they capture the "quiet" as well as AR coronal plasma that is ejected with  the CME. As solar flare emission which might hamper the identification of coronal dimmings is more pronounced in 193\,\AA, we use for the coronal dimming analysis the 211\,\AA\ filter.

Using logarithmic base-ratio images, constructed by dividing each image in the time series under study by a set of pre-event ``base images'', we track coronal dimmings using the threshold-based method of \cite{dissauer+2018a_apj,dissauer+2018b_apj}, which was further developed for the application to dimmings observed off-limb by \cite{2020ApJ...896...17C}. The cumulative dimming area at time $t$, $A(t)$, is defined by the sum of all pixels that have been flagged as dimming pixels up to time $t$. Its derivative, $dA/dt$, represents the dimming area growth rate, i.e. shows how fast the dimming is growing over time. We estimate the total cumulative dimming flux using HMI LOS magnetograms within the ``magnetic dimming region'', defined as to where the flux density exceeds $10$\,\gauss\ \cite[cf.\,][]{dissauer+2018a_apj}. Analogously as for the flare ribbon analysis, we apply a $\pm$5\% change to the threshold level, and compute the mean values and corresponding standard deviation of the reconnection flux for further analysis.

\subsection{(I)CME reconstruction} 
\label{ss:methods_icme}

The CME related to the flaring region is well observed by SOHO/LASCO \citep{1995SoPh..162..357B} and STEREO-A Sun Earth Connection Coronal and Heliospheric Investigation \citep[SECCHI;][]{howard08} in the white-light coronagraph data (for a visualization of the relative position of operating satellites see Section\,\href{ss:ip_n_insitu}{\ref{ss:ip_n_insitu}}). Having two vantage points on the well-developed CME, we apply the graduated cylindrical shell model \citep[GCS;][]{2006ApJ...652..763T,2009SoPh..256..111T}. GCS is a 3D reconstruction technique to derive the CME geometry, propagation direction, and 3D (de-projected bulk) speed by fitting a projection of a 3D croissant geometry on 2D images from at least 2 different vantage points simultaneously. To follow the CME evolution through interplanetary space and to clearly link it to the {\it in-situ} measurements, the GCS results (propagation direction, speed, half-angle, tilt with respect to ecliptic) are fed into the drag based ensemble model \citep[DBEM; see][]{2013SoPh..285..295V,2018ApJ...854..180D,2021SoPh..296..114C}. DBEM is a CME propagation tool and gives probabilistic predictions of the CME arrival time and speed at a given target in the solar system. The model results are used to compare with \textit{in-situ} observations in order to unambiguously link between CME signatures observed close to the Sun and in interplanetary space. The \href{https://swe.ssa.esa.int/graz-dbem-federated}{DBEM} is available as ESA (European Space Agency) space weather service from the Heliospheric Weather Expert Service Centre (H-ESC).

\subsection{In-situ analysis} 
\label{ss:methods_insitu}

We analyze \textit{in-situ} signatures of the ICME at {\it Solar Orbiter} \citep[][]{muller20} and near-Earth using data from the \href{https://omniweb.gsfc.nasa.gov/html/ow_data.html#norm_pla}{OMNI} database \citep{king05}. To analyze magnetic field properties at {\it Solar Orbiter} we use 1-min averaged Level-2 data in RTN (Radial-Tangential-Normal; a spacecraft centered coordinate system) coordinates from the magnetometer \citep[MAG;][]{horbury20}, whereas for plasma properties we use Level-2 data with moments computed from the proton part of ion distribution function measured by the Solar Wind Analyser \citep[SWA;][]{owen20} Proton and Alpha particle Sensor (PAS), taken from the {\it Solar Orbiter} archive \href{http://soar.esac.esa.int/soar/}{(SOAR)}. SWA-PAS data were further processed in order to obtain 1-min averages. For the analysis of near-Earth magnetic and plasma properties, we use 1-min averaged OMNI magnetic field and plasma data in GSE (Geocentric Solar Ecliptic) coordinates, which are time-shifted to Earth's bow shock. In order to compare the orientation of the magnetic field near Earth and at {\it Solar Orbiter} we convert {\it Solar Orbiter} RTN coordinates to GSE-aligned coordinates by using the following substitutions: $B_r$\,$=$\,$-B_x$ and $B_t$\,$=$\,$-B_y$. In the RTN system the radial component is aligned with the Sun-spacecraft line and the tangential direction is defined through the projection of the solar rotational axis to the plane perpendicular to the radial component. In the GSE system, the $x$-component is aligned with the Earth-Sun line and the $z$ direction points towards the ecliptic north pole  \citep[for overview of coordinate systems see e.g.][]{2002P&SS...50..217F}. Therefore, for a spacecraft aligned perfectly radially with Earth and lying in the ecliptic plane, $B_r$\,$=$\,$-B_x$ and $B_t$\,$=$\,$-B_y$ are exact coordinate transformations. Given the position of {\it Solar Orbiter} at the considered time this substitution is a solid approximation for this event.

Uncertainties of the {\it in-situ}-based quantities are approximated based on the respective scales used for their visualization and subsequent analysis (\href{fig:insitu}{Figure~\ref{fig:insitu}} in Section~\href{ss:ip_n_insitu}{\ref{ss:ip_n_insitu}}). Therefore, we assume a quarter of the time scale (0.025~DOY) for the uncertainty in arrival times, 1\,nT  for the uncertainty of the magnetic field and 10\,\kmps for that of the flow speed. The uncertainties of other parameters are derived using rules of error propagation.

\begin{figure*}
\centering
\includegraphics[width=0.9\textwidth]{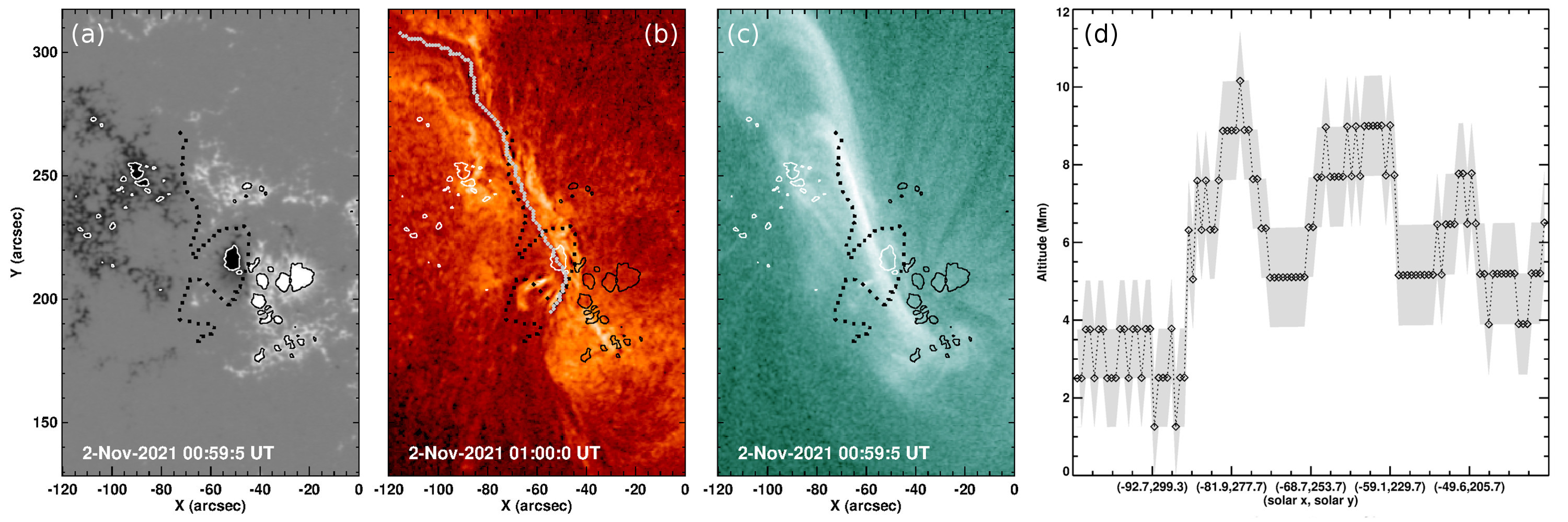}
\caption{Observations and stereoscopic reconstruction of AR~12891 on 2021~November~2 at 00:59\,UT. (a) Vertical photospheric magnetic field saturated at $\pm$1\,kG. Contours are drawn at $\pm$0.75\,kG. The main PIL is indicated by a black dotted curve. (b) Nearest-in-time AIA~304\,\AA\ unsharp-mask image. The path along of which stereoscopic reconstruction of the height of the filament is performed is indicated by gray crosses (differentially rotated from the time of stereoscopic reconstruction at 02~November 01:05\,UT). (c) Nearest-in-time AIA~94\,\AA\ image. (d) Stereoscopy-based estimate of the height of the filament (black diamonds) and associated uncertainty (gray-shaded area).
}
\label{fig:preflare_corona}
\end{figure*}

\section{Results}
\label{s:results}

\subsection{Pre-flare corona} 
\label{ss:structure}

The main polarity inversion line (PIL) that separates the two major polarities of AR~12891 is oriented roughly along of the solar north-south direction (black dotted curve in Figure\,\href{fig:preflare_corona}{\ref{fig:preflare_corona}(a)}). ARs with such a basic morphology \citep[dubbed "spot-spot" type by][see their Figure\,6]{2017ApJ...834...56T} may be created by many episodes of flux emergence and may produce flares with long elongated ribbons. The AR hosted a total unsigned magnetic flux of a few $10^{21}$\,\mx, placing it at the lower end of corresponding distributions \citep[e.g., Figures~ 8 and 9 of][]{2017ApJ...845...49K}. Spatially associated with the main PIL, a pronounced filament is observed in the form of a dark elongated structure in AIA~304\,\AA\ images (Figure\,\href{fig:preflare_corona}{\ref{fig:preflare_corona}(b)}). We apply stereoscopic means to provide a purely observation-based estimate of the coronal height of the filament, where we find heights of $\lesssim$10\,Mm (Figure\,\href{fig:preflare_corona}{\ref{fig:preflare_corona}(d)}). At higher temperatures bright emission in the form of inverse-S-shaped coronal loops is observed, indicating a left-handed underlying magnetic field (Figure\,\href{fig:preflare_corona}{\ref{fig:preflare_corona}(c)}). The stereoscopic estimates for three time instances for which data was available to be analyzed (01~November 23:05\,UT, 02~November 00:05\,UT and 01:05\,UT; not shown explicitly) indicate that the height of the filament channel remained more or less constant.

\begin{figure*}
\centering
\includegraphics[width=0.9\textwidth]{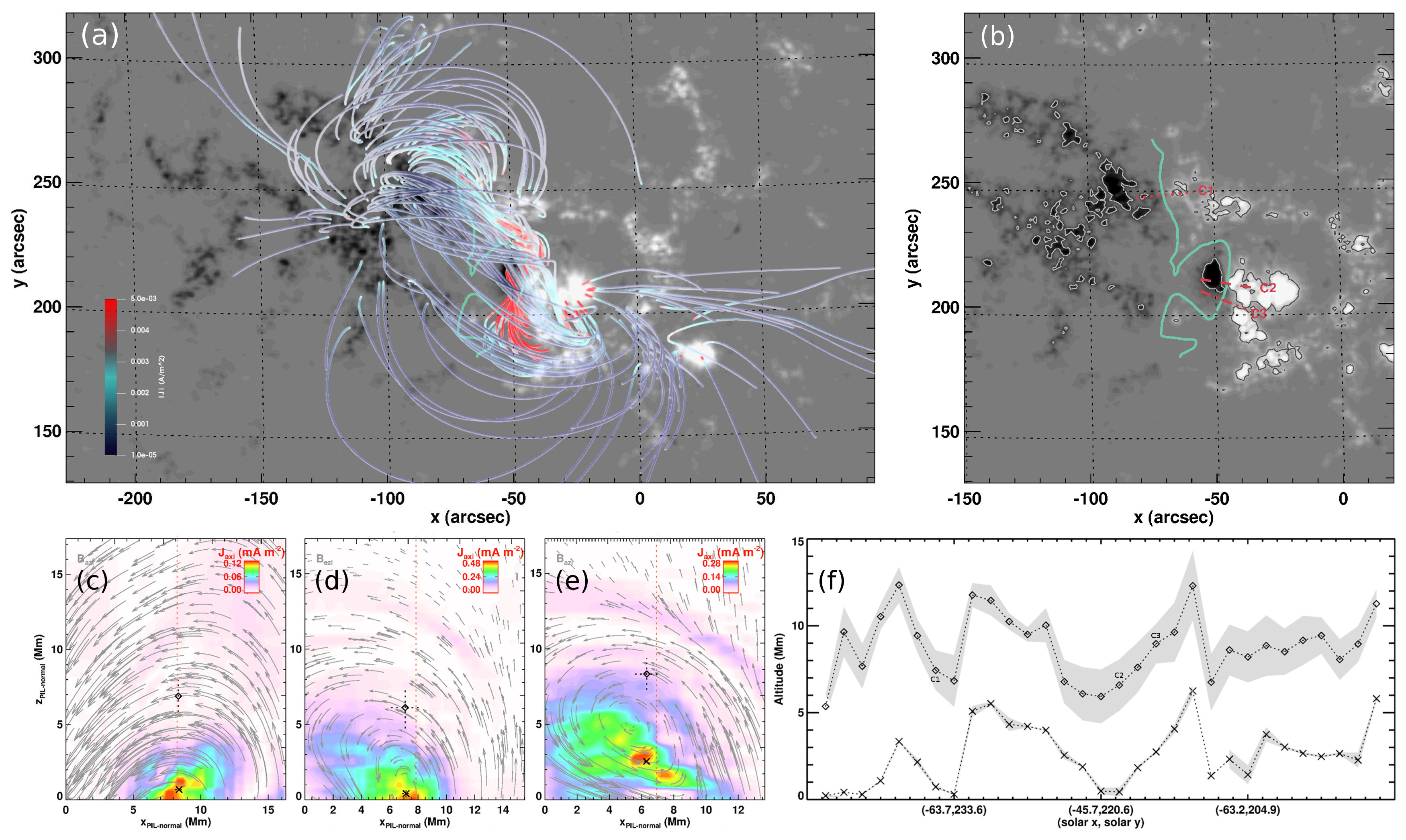}
\caption{Coronal magnetic field modeling of AR~12891 on 2021~November~2 at 00:59\,UT. (a) Field line connectivity. Model field lines are drawn from randomly selected footpoints and color-coded according to the magnitude of the local electric current density. The gray-scale background resembles the vertical photospheric magnetic field component, saturated at $\pm$1\,kG. (b) Vertical component of photospheric magnetic field saturated at $\pm$1\,kG. Black and white contours are drawn at $\pm$0.75\,kG. The main PIL is indicated by the green curve. Red straight lines resemble the footprints of selected vertical slices, labeled C1--C3, for which the spatial distribution of the azimuthal magnetic field (arrows) and unsigned axial electric current density (color-coded background) is shown in panels (c)--(e), respectively. Estimates of the average height of the center of the flux rope and arcade field within the selected slices are indicated as crosses and diamonds, respectively. Dashed lines indicate corresponding uncertainties. (f) Estimate of the altitude of the flux rope center (plus signs) and envelope (diamonds) along of the main PIL, based on 40 vertical slices distributed regularly along of the PIL.
}
\label{fig:b_field}
\end{figure*}

NLFF modeling of the coronal magnetic field in and around AR~12891 at 2~November 00:59\,UT (Figure\,\href{fig:b_field}{\ref{fig:b_field}(a)}) reveals a pronounced coronal arcade extending up to $\gtrsim$60\,Mm (blueish color), overlying strongly twisted field (reddish color), the latter being spatially associated to the main PIL (Figure\,\href{fig:b_field}{\ref{fig:b_field}(b)}) as well as to the the filament observed in AIA~304\,\AA\ (see Figure\,\href{fig:preflare_corona}{\ref{fig:preflare_corona}(b)}). The strongly twisted model magnetic field exhibits a rotational pattern around a common axis and is therefore referred to as a flux rope in the following. When viewed along of the solar south-north direction (i.e., along of the field starting from the positive-polarity area in the south-west of the AR), it exhibits a left-handed rotation, indicative of a dominant left-handed helicity of the active-region coronal field.

From projection of the 3D NLFF magnetic field into vertical planes oriented locally perpendicular to the main PIL (see Figure\,\href{fig:b_field}{\ref{fig:b_field}(c)}--\href{fig:b_field}{\ref{fig:b_field}(e)} for a visualization and their footprints labeled C1--C3 in Figure\,\href{fig:b_field}{\ref{fig:b_field}(b)}, respectively) we approximate the height of the flux rope along of the PIL (green line in Figure\,\href{fig:b_field}{\ref{fig:b_field}(b)}). More precisely, we use the unsigned axial current density ($J_{\rm axi}$; see color-coded background in Figure\,\href{fig:b_field}{\ref{fig:b_field}(c)}--\href{fig:b_field}{\ref{fig:b_field}(e)}) to estimate the average height of the flux rope center (indicated by crosses) and of the associated arcade field (indicated by triangles). The averages are computed using selected thresholds in the regimes $>$\,0.75\,$\times$\,$J_{\rm axi,max}$ (assuming that the strongest currents are located near the central axis) and $<$\,0.25\,$\times$\,$J_{\rm axi,max}$ (assuming a rapid decay of $J_{\rm axi}$ towards the arcade field). We find from the NLFF-model based estimate that the center of the flux rope (where the strongest electric currents reside) is located at altitudes of $\lesssim$5\,Mm above the NLFF model's lower boundary (represented by crosses in Figure\,\href{fig:b_field}{\ref{fig:b_field}(f)}). The associated arcade field extends up to altitudes in the approximate range of 5--10\,Mm (represented by diamonds), in close agreement with the stereoscopic estimate. 

From the 22 NLFF models employed for the considered time interval, we estimate the mean axial flux along of the flux rope as $\approx$\,2.9$\pm$1.7$\times10^{20}$\,\mx\ and the average magnetic field as $\approx$\,262$\pm$126\,\gauss. Furthermore, from the NLFF model at 02~November 00:59\,UT, we estimate a free magnetic energy of $\simeq$\,5.7\,$\times10^{32}$\,\erg\ and a current-carrying helicity of $\simeq$\,$-$3.2\,$\times10^{41}$\,\mxmx, the latter being consistent with the left-handed sense of the model flux rope. The time evolution of the coronal magnetic energies and helicities is analyzed in detail in Section\,\ref{ss:activity} and interpreted in context with observations-based measures of flare-related emission. 

\begin{figure*}
\centering
\includegraphics[width=0.9\textwidth]{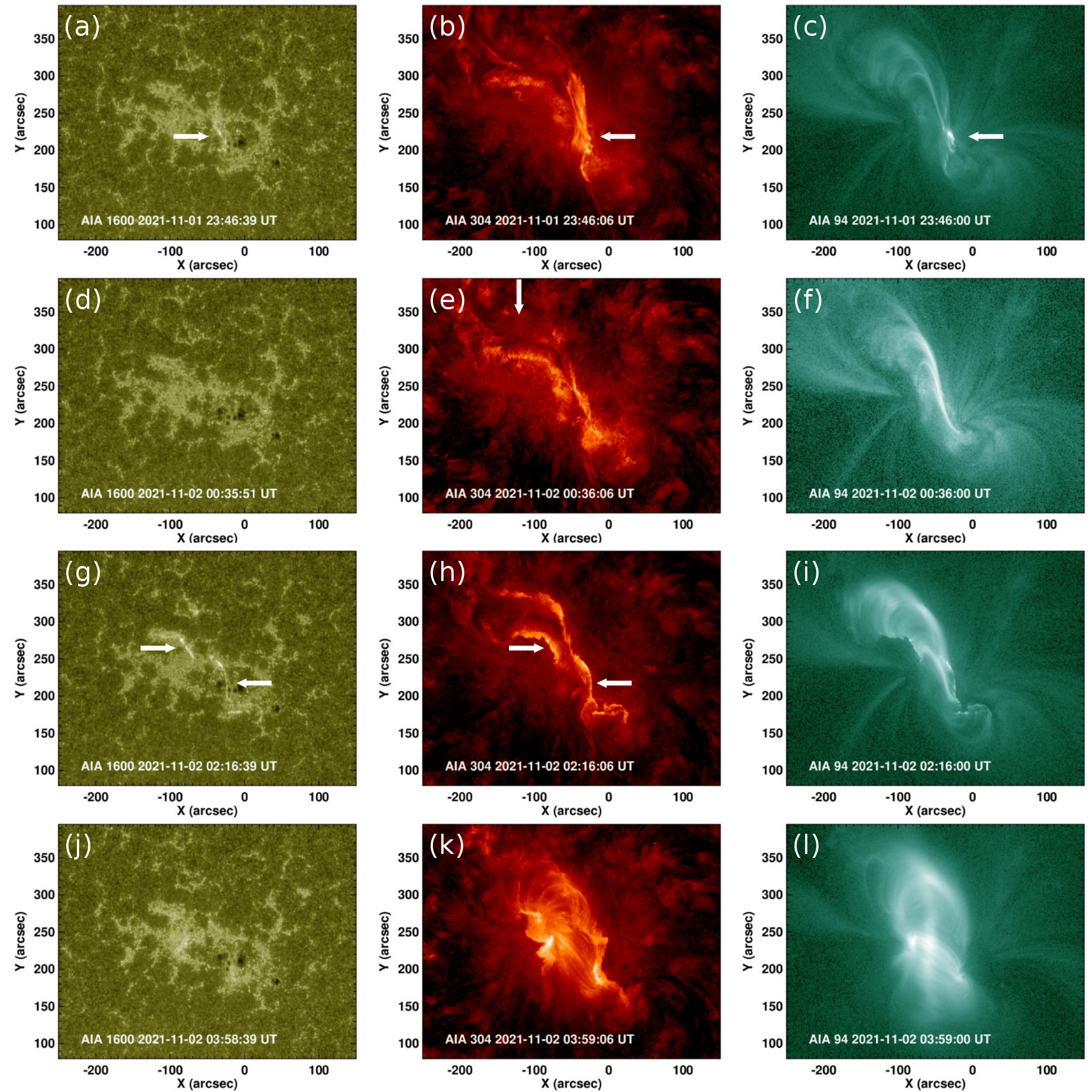}
\caption{EUV images showing the main features of activity in AR~12891 between late November~01 and early November~02: A C4.5 flare (\textit{top row}), a subsequent partial filament eruption (\textit{2nd row}), the early stages of a double-peak M-class flare (\textit{3rd row}), as well as its aftermath (\textit{bottom row}). From left to right, AIA 1600\,\AA, 304\,\AA, and 94\,\AA\ filtergrams are shown. A movie accompanying the figure is available in the electronic material (flare.mp4).}
\label{fig:sdo_still}
\end{figure*}

\begin{figure*}
\centering
\includegraphics[width=0.9\textwidth]{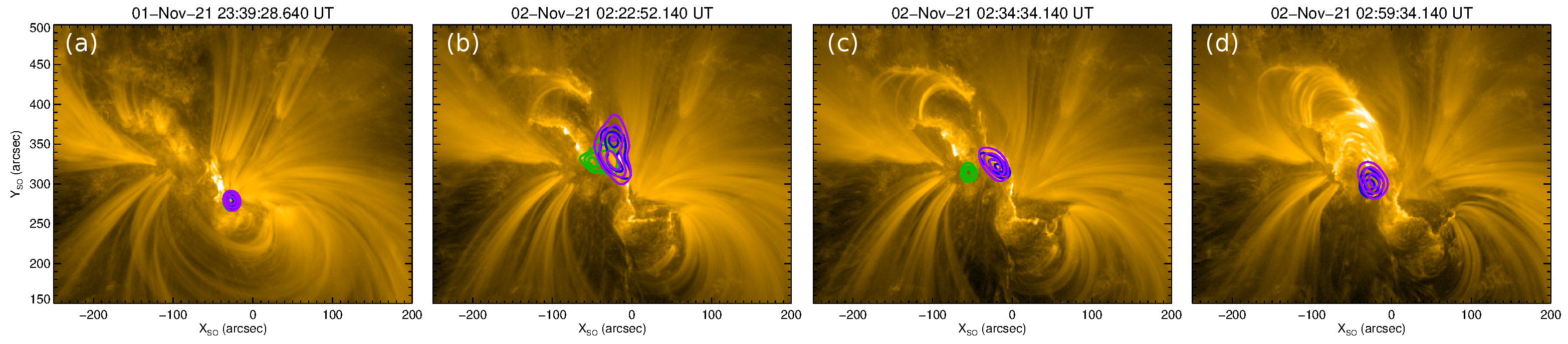}
\caption{STIX images showing 4--1\,\keV (thermal; purple), 11--16\,\keV (intermediate; blue) and 16--28\,\keV (non-thermal; green) contours, at the 50\%, 70\% and 90\% levels, each integrated over a 2-min period centered around a peak in the STIX 16--28\,\keV energy band (cf.\ Figure\,\href{fig:model_vs_obs}{\ref{fig:model_vs_obs}(b))}. The STIX images were generated using the Maximum Entropy Method MEM\_GE \citep{Massa_2020} with an AIA 171\,\AA\ base selected from around the mean time of the image. Units are arcseconds within the {\it Solar Orbiter} view. The AIA images have been rotated to the {\it Solar Orbiter} viewpoint.}
\label{fig:aia_stix}
\end{figure*}

\subsection{Eruptive activity}
\label{ss:activity}

Figure\,\href{fig:sdo_still}{\ref{fig:sdo_still}} shows the EUV emission of the AR under study between 01~November $\sim$23:46\,UT and 02~November $\sim$04:00\,UT, at selected time instances representative for the main eruptive activity: a C4.5 flare that peaked at 01~November $\sim$23:40\,UT, a partial filament eruption timely centered around 02~November $\sim$00:30\,UT, and the M1.7 flare that peaked at 02~November $\sim$03:01\,UT. In the following, these observations (see corresponding movie for clarity) are interpreted in context with spatial and temporal aspects of simultaneously detected coronal dimmings, the latter interpreted as to represent signatures that stem from the evacuation of plasma material when a CME expands outwards. 

\begin{figure*}
\centering
\includegraphics[width=0.85\textwidth]{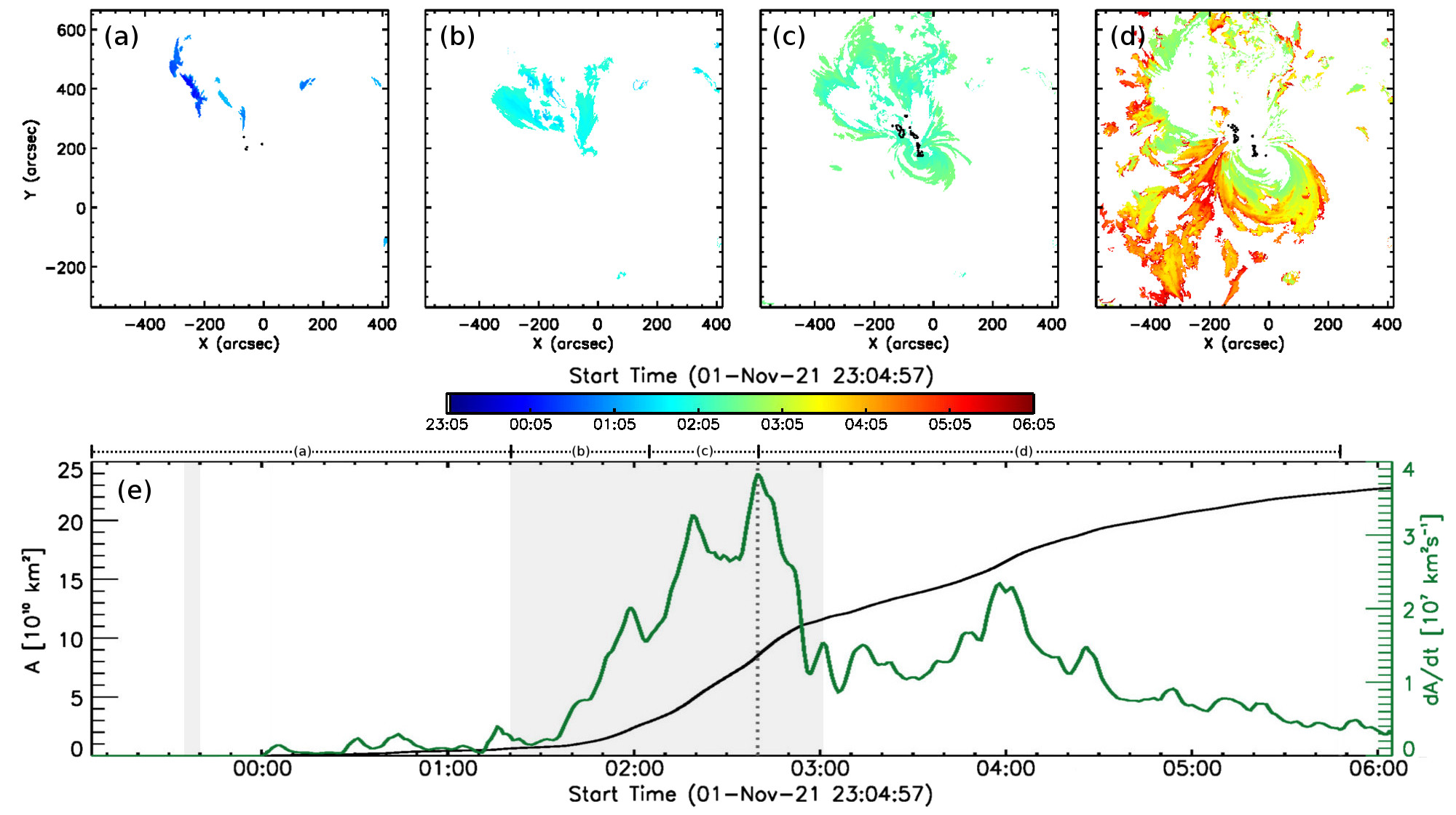}
\caption{Spatial and temporal evolution of coronal dimmings and flare ribbons. {\it Top:} Area newly occupied by coronal dimmings (color-coded according to time) during individual episodes: (a) the C4.5 flare and a subsequent (flare-less) partial filament eruption (01~November 23:05 -- 02~November 01:20\,UT), (b) the early phase of the M1.6 flare (01:20 -- 02:05\,UT), (c) the impulsive phase of the M1.6 flare (02:05 -- 02:40\,UT), and (d) the early decay phase of the M1.7 flare (02:40 -- 05:48\,UT). The corresponding total area occupied by flare ribbons is outlined as black contour. {\it Bottom:} Cumulative dimming area (black) and instantaneous growth rate (green) as a function of time. The width of the time windows covered in (a) -- (d) is indicated by dotted horizontal lines at the top axis in (e). Gray-shaded vertical bands mark the impulsive phases of flares.}
\label{fig:dtiming}
\end{figure*}

During the C4.5 flare late on November~01, part of the observed filament erupted (Figure\,\href{fig:sdo_still}{\ref{fig:sdo_still}(b)}). Correspondingly, distinct flare kernels appear to the east of the leading sunspot (Figure\,\href{fig:sdo_still}{\ref{fig:sdo_still}(a)}) that developed into short northward progressing flare ribbons, along with the appearance of pronounced intermediate and non-thermal X-ray sources (Figure\,\href{fig:aia_stix}{\ref{fig:aia_stix}(a)}). Simultaneous EUV observations reveal enhanced emission along of loop-like structures connecting to the northern parts of the AR (Figure\,\href{fig:sdo_still}{\ref{fig:sdo_still}(c)}) yet without signatures characteristic for the successful expulsion of coronal plasma in the form of a non-zero dimming growth rate (green curve in Figure\,\href{fig:dtiming}{\ref{fig:dtiming}(e)}). Later, around 02~November $\sim$00:35\,UT the northernmost segments of the filament channel erupted (visible as outward moving dark features in AIA~304\,\AA\ maps; indicated by the white arrow in Figure\,\href{fig:sdo_still}{\ref{fig:sdo_still}(e)}; see also accompanying movie for clarity) and accompanied by first spatially pronounced dimming features (see non-zero dimming area and growth rate in Figure\,\href{fig:dtiming}{\ref{fig:dtiming}(a)} and \href{fig:dtiming}{\ref{fig:dtiming}(e)}, respectively). 

\begin{figure*}
\centering
\includegraphics[width=0.85\textwidth]{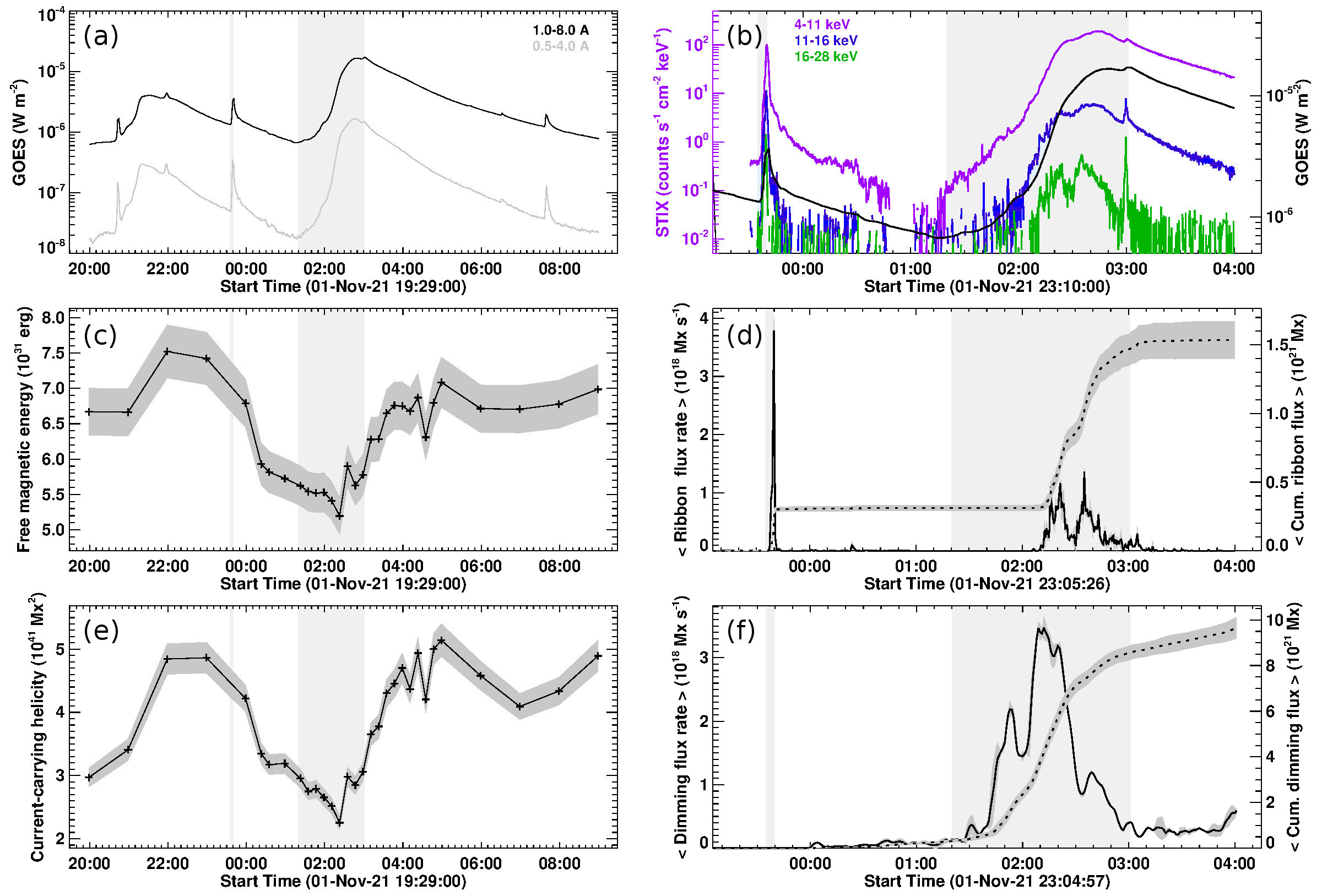}
\caption{Time evolution of (a) GOES 1--8\,\AA\ (black) and 0.5--4\,\AA\ (gray) SXR flux. (b) STIX count rates at 4--11\,keV (purple), 11--16\,keV (blue), and 16--28\,keV (green) energies together with the GOES 1--8\,\AA\ SXR flux (black). (c) Free magnetic energy. (d) Reconnection flux change rate (solid) and cumulative reconnection flux (dashed) in flare ribbons. (e) Magnitude of the helicity of the current-carrying field ($|\Hj|$). (f) Reconnection flux change rate (solid) and cumulative reconnection flux (dashed) in dimmings. Gray-shaded vertical bands mark the impulsive phases of flares. The panels in the right column cover the time around the eruptive activity (01~November 23:00\,UT -- 02~November 04:00\,UT) while the panels in the left column cover an extended time range (01~November 20:00\,UT -- 02~November 09:00\,UT).}
\label{fig:model_vs_obs}
\end{figure*}

During the early phase of the M1.6 flare, extended flare ribbons developed after $\sim$02:14\,UT (Figure\,\href{fig:sdo_still}{\ref{fig:sdo_still}(g)}--\href{fig:sdo_still}{\ref{fig:sdo_still}(h)}) yet lacking signatures of an obviously erupting structure. The latter appeared only after $\sim$02:20\,UT and was directed towards solar south (see accompanying movie). At that time pronounced X-ray sources developed too (Figure\,\href{fig:aia_stix}{\ref{fig:aia_stix}(b)}--\href{fig:aia_stix}{\ref{fig:aia_stix}(c)}) and the maximum growth rate in dimming area was reached ($\approx$\,4$\times10^7$\,\kmsps; see vertical dotted line in Figure\,\href{fig:dtiming}{\ref{fig:dtiming}(e)}), the latter associated to the growth of dimming area both towards the solar north and south direction (Figure\,\href{fig:dtiming}{\ref{fig:dtiming}(c)}). A broad bright post-flare arcade, spanning the whole underlying filament channel appears in EUV after $\sim$03:00\,UT (Figure\,\href{fig:sdo_still}{\ref{fig:sdo_still}(k)}--\href{fig:sdo_still}{\ref{fig:sdo_still}(l)}), along with mainly thermal X-ray sources (Figure\,\href{fig:aia_stix}{\ref{fig:aia_stix}(d)}) and along with pronounced dimmings mainly towards the solar south direction (Figure\,\href{fig:dtiming}{\ref{fig:dtiming}(d)}).

The temporal profiles of the full-disk GOES 1--8\,\AA\ and 0.5--4\,\AA\ SXR flux (Figure\,\href{fig:model_vs_obs}{\ref{fig:model_vs_obs}(a)}) show a short-lived enhanced emission associated to the C4.5 flare late on November~01 as well as a long-duration enhanced emission associated to the M-class flaring early on November~02. A closer inspection of the temporal profile reveals that the latter actually consisted of two main episodes: a preceding prolonged one which started at $\sim$01:20\,UT and peaked around $\sim$02:51\,UT (an M1.6-class flare) as well as a subsequent narrow peak at M1.7 level at $\sim$03:01\,UT. For completeness we note that the C-flare activity before the C4.5 flare was related to other sources on the solar disk.

The time evolution of the coronal free energy, $\Ef$, during that period of enhanced flare activity of AR~12891 was characterized by an overall decrease from values of $\gtrsim$\,7\,$\times$\,$10^{31}$\,\erg\ late on November~01 to $\lesssim$\,6\,$\times$\,$10^{31}$\,\erg\ prior to the M-class flaring early on November~02, followed by an increase back to a nearly pre-C-flare level in its aftermath (Figure\,\href{fig:model_vs_obs}{\ref{fig:model_vs_obs}(c)}). In a similar manner, the magnitude of the helicity of the current-carrying field, $\Hj$, decreases from values $\gtrsim$\,4.5\,$\times$\,$10^{41}$\,\mxmx\ prior to the C-class flare to values of $\lesssim$\,3.5\,$\times$\,$10^{41}$\,\mxmx\ prior to the M-class flaring and is replenished back to an approximate pre-C-flare level quickly afterwards (within $\sim$two hours; Figure\,\href{fig:model_vs_obs}{\ref{fig:model_vs_obs}(e)}). Notably, pronounced decreases of the coronal free energy and current-carrying helicity occur before 02~November $\approx$\,02:40\,UT, the time when the dimming area growth rate peaked (compare Figure\,\href{fig:dtiming}{\ref{fig:dtiming}(e)}). Considering mean "pre-flare" (in the time range 01~November 22:00--23:00\,UT) and post-flare (in the time range 02~November 03:00--04:00\,UT) values, the notable decreases in the coronal budgets amount to $\Delta\Ef$\,$\approx$\,1.4$\pm$0.4$\times10^{31}$\,\erg\ and $\Delta|\Hj|$\,$\approx$\,1.5$\pm$0.4$\times10^{41}$\,\mxmx, supporting the scenario of a successful ejection of left-handed (negative-helicity) magnetic field.

Inspection of eruptivity-related emission allows us to establish a link to the magnetic flux involved, i.e., processed by means of magnetic reconnection. On the one hand, flare ribbons stem from the energy deposition of flare-accelerated electrons precipitating downward along of newly reconnected field. The electron beams also cause the non-thermal hard X-ray emission spatially associated to the low-atmosphere where the newly reconnected field is anchored (see Figure\,\href{fig:aia_stix}{\ref{fig:aia_stix}}). On the other hand, coronal dimmings reflect plasma evacuation in the low corona. The time evolution of the mean magnetic flux change rates associated to flare ribbons and dimmings (both on the order of $10^{18}$\,\mxps during the times eruptive activity) are shown in Figure\,\href{fig:model_vs_obs}{\ref{fig:model_vs_obs}(d)} and \href{fig:model_vs_obs}{\ref{fig:model_vs_obs}(f)}, respectively, and are summarized in context with the observed X-ray emission in the following.

The temporal profile of the STIX 4--11\,\keV\ (thermal) and 11--16\,\keV\ (intermediate) X-ray emission appears similar to that of the GOES SXR flux, yet exhibiting much more detail. Besides enhanced emission co-temporal with the GOES SXR peak times of the C4.5 (at 01~November 23:40\,UT), M1.6 (at 02~November 02:51\,UT) and M1.7 (at 02~November 03:01\,UT) flares, two additional peaks are noticed in the intermediate and non-thermal (16--28\,\keV) energy bands centered around $\sim$02:20\,UT and $\sim$02:40\,UT (see red and green curves in Figure\,\href{fig:model_vs_obs}{\ref{fig:model_vs_obs}(b)}). Noteworthy, the latter two are observed co-temporal with peaks in the dimming growth rate (compare green curve in Figure\,\href{fig:dtiming}{\ref{fig:dtiming}(e)}) and to two pronounced peaks in the mean flare ribbon flux change rate (compare black solid curve in Figure\,\href{fig:model_vs_obs}{\ref{fig:model_vs_obs}(d)}). The latter also exhibits a strongest peak co-temporal with the impulsive phase of the C4.5-class flare. In comparison, the mean magnetic flux change rate in coronal dimmings (black solid curve in Figure\,\href{fig:model_vs_obs}{\ref{fig:model_vs_obs}(f)}) exhibits no obvious response to the C4.5-class flare and strongest responses occur during the early impulsive phase of the M1.6 flare, i.e., before notable ribbon-associated fluxes are detected. From the mean cumulative flare ribbon flux (black dashed curve in Figure\,\href{fig:model_vs_obs}{\ref{fig:model_vs_obs}(d)}), we estimate that a total of $\approx$\,$1.5\pm0.1$\,$\times$\,$10^{21}$\,\mx\ was liberated between 01~November 23:40\,UT and 02~November 04:00\,UT, i.e., during the course of the C4.5 and M-class flares. Similarly, from the cumulative dimming flux (black dashed curve in Figure\,\href{fig:model_vs_obs}{\ref{fig:model_vs_obs}(f)}), we estimate a total of $\approx$\,$9.7\pm0.5$\,$\times$\,$10^{21}$\,\mx.

\begin{figure*}
\centering
\includegraphics[width=0.9\textwidth]{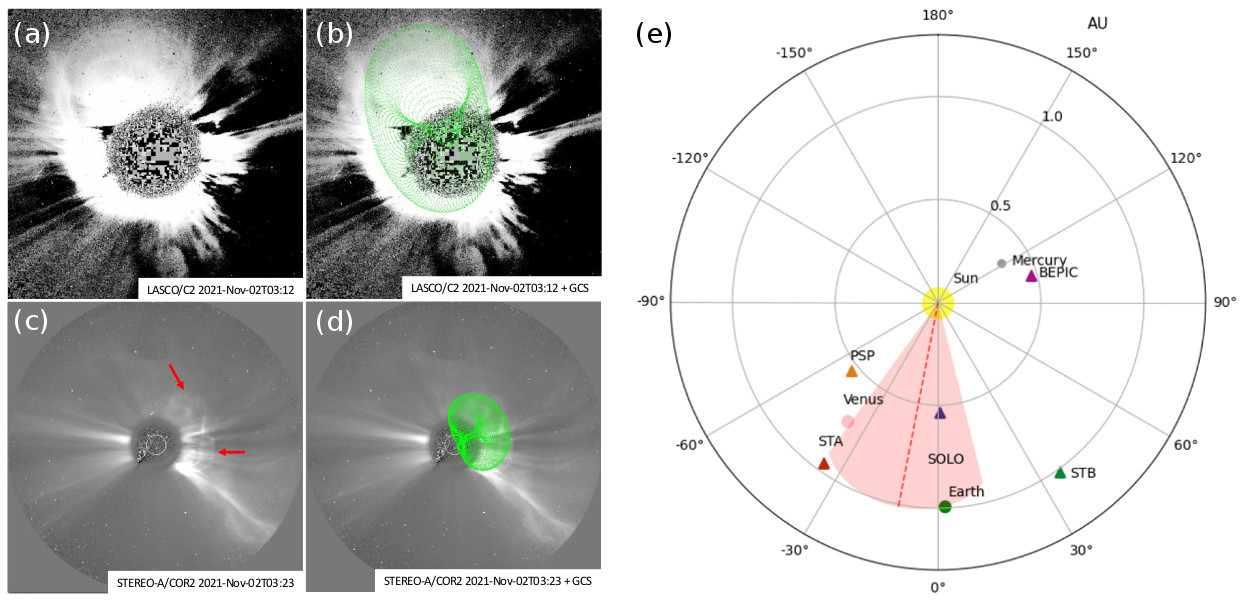}
\caption{Observations and modeling of the (I)CME. Observations of the CME as seen from (a) LASCO/C2  and (c) STEREO-A. Bright emission fronts seen in STEREO-A are indicated exemplary by red arrows. GCS-reconstructed CME fronts are shown as green meshs in (b) and (d), respectively, on the respective white-light coronagraph data. (e) DBEM-based CME propagation direction (red dashed line) and width (red-shaded area) in context with the interplanetary position of operating spacecrafts (triangles) and planets (bullets).}
\label{fig:icme}
\end{figure*}

\subsection{Upper corona and interplanetary space}
\label{ss:ip_n_insitu}

In LASCO imagery the CME appears as a halo event directed to the North-East (Figure\,\href{fig:icme}{\ref{fig:icme}(a)}). STEREO-A, located at $\sim$34~degrees East of the Sun-Earth line, observes the CME from a side-view as it propagates away from the North-West and South-West quadrants (Figure\,\href{fig:icme}{\ref{fig:icme}(c)}). In the coronagraph data, a CME appears first in LASCO/C2 at 01~November $\sim$02:00\,UT, in STEREO-A/COR1 at $\sim$01:31\,UT, and in STEREO-A/COR2 at $\sim$01:53\,UT.  From visual inspection of images at later times, i.e., when the CME has developed further into interplanetary space, we identify multiple fronts (see red arrows in Figure\,\href{fig:icme}{\ref{fig:icme}(c)}) hinting at multiple (at least two) eruptions: an earlier one oriented towards the North-West and a subsequent one heading towards the South-West. 

We reconstruct the flux rope geometry of the ICME using the GCS method applied to the coronagraph white-light image data from LASCO/C2 on 02~November 03:12\,UT and from STEREO-A/COR2 on 02~November 03:23\,UT (see green meshes in Figure\,\href{fig:icme}{\ref{fig:icme}(b)} and \href{fig:icme}{\ref{fig:icme}(d)}, respectively). The GCS parameters longitude/latitude are derived as E10/N20, half-width $\alpha$\,$=$\,25~degrees \citep[calculated using the relation given in][]{2019ApJ...880...18D}, aspect ratio $\kappa$\,$=$\,0.35\,rad and tilt angle as $-75$~degrees. To calculate the 3D speed, we derive the CME height from subsequent STEREO-A/COR2 and LASCO/C3 images, covering the time period up to 04:23\,UT, and assuming a self similar expansion. For that purpose we keep constant the GCS-based parameters $\alpha$, $\kappa$, longitude, latitude, and tilt angle, and only vary the height. As a result we find that on 02~November at 04:23\,UT, the CME apex reached a height of 17.5 solar radii (Rs) with an average speed of 1600\,\kmps. 

The GCS-derived CME parameters (speed and angular half-width at a certain time and distance, as well as the longitude of the solar source region) and their default uncertainty ranges (time $\pm$30 minutes;  angular half-width $\pm$15~degrees;  speed $\pm$200\,\kmps; longitude $\pm$30~degrees) together with the a priori unknown values of the drag parameter ($\gamma$) and ambient solar wind speed ($u$) are used as input for the DBEM in order to connect the ICME signatures as observed close to the Sun to those measured {\it in-situ}. The unknown values of $\gamma$ and $u$ are chosen such that the modeled ICME mean arrival time and speed for {\it Solar Orbiter} and Earth is  matching well with the observed arrival times and speeds at the targets (we allowed for a difference in the arrival times of maximum 2~hours and of maximum 50\,\kmps\ in speed). Varying values as $\gamma$\,=\,(0.25$\pm$0.1)$\times10^{-7}$\,km$^{-1}$ and $u$\,=\,(500$\pm$50)\,\kmps, the best agreement with the observed ICME arrival time and speed is found for Earth (with a difference of only a few minutes between the modeled and observed values; see Table~\ref{tab:dbem}). For {\it Solar Orbiter} the estimated arrival time of the ICME is $\approx$\,1.5\,hours too early with a difference in speed of about 25\,\kmps. Those values, however, are clearly within the statistical uncertainties \citep[see][]{2013SoPh..285..295V}, hence, supporting the connection between CME structures observed close to the Sun and the {\it in-situ} measured signals. With the same input parameters we also run DBEM for the target STEREO-A and compare the results to the \textit{in-situ} measurements. For STEREO-A the ICME is predicted to arrive about 3.5~hours too early with a difference between modeled and observed speed of about 120\,\kmps. Figure\,\href{fig:icme}{\ref{fig:icme}(e)} illustrates the results from the DBEM simulation, showing the CME propagation direction (red dashed line) and width (red-shaded area) together with the constellation of planets and spacecrafts in operation. The DBEM result reveals that {\it Solar Orbiter} and near-Earth spacecraft are located close to the ICME apex while STEREO-A is hit by its flank only, the latter explaining the larger differences between modeled and observed arrival time and speed. The spatial information derived from DBEM supports the analysis and interpretation of the {\it in-situ} measurements by {\it Solar Orbiter} and the ACE/Wind satellites in the following. 

\begin{table}[t]
\footnotesize
\caption{
\label{tab:dbem}
{\it In-situ} measured/observed (O) CME arrival time and arrival speed at several targets ({\it Solar Orbiter}, Earth, and STEREO-A), compared to the DBEM propagation model results (M).}
\centering
\begin{tabular}{|l|l|l|}
\hline
  target  & arrival time M/O [DOY@UT]  & $v \binom{\rm max}{\rm min}$ M/O [km/s] \\
\hline
    {\it Solar}  & \multirow{2}{*}{307@13:00($\pm$3.5h)/307@14:24} & \multirow{2}{*}{755$\binom{848}{682}$/730 }\\
    {\it Orbiter} & & \\
Earth & 307@21:40($\pm$4.0h)/307@21:36 & 712$\binom{795}{649}$/710 \\
STEREO-A & 308@00:28($\pm$3.5h)/308@04:00 & 678$\binom{791}{647}$/550 \\
\hline
\end{tabular}
\normalsize
\end{table}

\begin{figure*}
\centering
\includegraphics[width=0.9\textwidth]{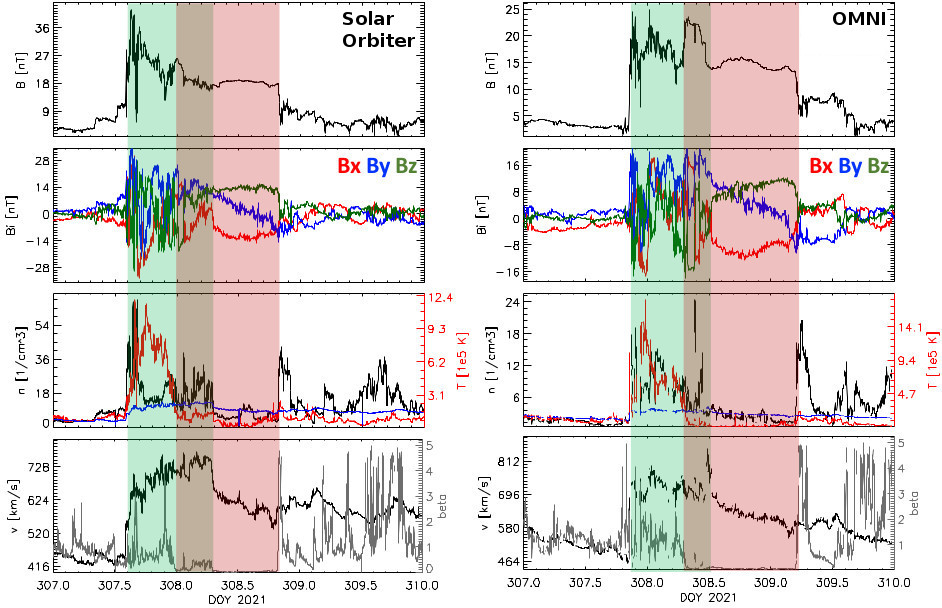}
\caption{
\textit{In-situ} measurements given in day-of-year (DOY) time series around 2021 November~04 (DOY 308) from {\it Solar Orbiter} (\textit{left panels}) and OMNI (\textit{right panels}). \textit{Top row:} Magnetic field strength. \textit{2nd row}: $x-$, $y-$, and $z-$component (red, blue and green, respectively) of the magnetic field in the Geocentric solar ecliptic (GSE). \textit{Third row}: Plasma density (black), plasma temperature (red) and expected temperature (blue). \textit{Bottom row}: Plasma flow speed (black) and plasma-$\beta$ (gray). Observed features corresponding to the sheath region, frontal region of the interplanetary flux rope and magnetic cloud are indicated by green-, olive- and red-shaded areas, respectively.}
\label{fig:insitu}
\end{figure*}

{\it Solar Orbiter} and OMNI {\it in-situ} data show great similarity (Figure\,\ref{fig:insitu}). In both we observe clearly a shock arrival, followed by characteristic sheath properties (green-shaded area), disturbed frontal region of a flux rope (olive-shaded) and MC signatures (red-shaded). We observe the shock arrival at {\it Solar Orbiter} at DOY~307.6 (01~November 14:30 UT) and at Earth at DOY~307.9 (03~November 21:30 UT), followed by a region of increased density and temperature, with high plasma beta and fluctuating magnetic field indicative of the arrival of the ICME body. Especially the magnetic field components at {\it Solar Orbiter} and Earth show a striking resemblance: starting at DOY~308.3 at {\it Solar Orbiter} (04~November 07:00 UT) and DOY~308.55 at Earth (04~November 13:00 UT) clear MC signatures are recorded \citep[for an overview on MC properties see, e.g.,][]{1982JGR....87..613K,2006SSRv..123...31Z,2017LRSP...14....5K}. These include enhanced magnetic field (top panels in Figure\,\ref{fig:insitu}), low proton temperatures (red curve in third row of Figure\,\ref{fig:insitu}) and low plasma-$\beta$ (gray curve in bottom panels of Figure\,\ref{fig:insitu}). Notably, throughout the MC, the $\By$ component of the magnetic field is rotating from positive to negative values, while the $\Bz$ component remains positive. This is indicative for a left-handed flux rope, highly inclined with respect to the ecliptic. According to the classification by \cite{bothmer98}, this is a east-north-west (ENW)-type of flux rope. Though not shown explicitly, from both {\it Solar Orbiter} and OMNI data, the polar angle of the magnetic field is positive (indicating north), whereas the azimuthal angle of the magnetic field rotates by roughly 140~degrees from the eastern direction to the western direction. More precisely, at {\it Solar Orbiter} we observe a rotation from $\sim$\,125 to $\sim$\,265~degrees, and in OMNI data from roughly 135 to 273~degrees. This, according to the classification by \cite{2019SoPh..294...89N}, corresponds to a flux rope with a single rotation in the range 90--180~degrees (``F\_r'') at both spacecraft.

We performed measurements of basic properties of the sheath, frontal region and MC at {\it Solar Orbiter} and Earth (see Table \ref{tab:in_situ}), and focus on the MC properties, as they can be compared to corresponding estimates from the solar source region. The MC shows a clear and symmetric profile, which can be approximated by a linear fit, indicating that a simple circular-cross-section Lundquist type of model is applicable \citep[see e.g.][]{demoulin19}. In particular, we are interested in the size, average magnetic field strength, axial magnetic flux, and helicity at {\it Solar Orbiter} and Earth. We assume that the ICME expands self-similarly in the radial direction from {\it Solar Orbiter} to Earth. We base this assumption on the fact that both at {\it Solar Orbiter} and Earth the flow speed exhibits a globally monotonically decreasing profile. This assumption is further supported by the visual similarity of the magnetic field configuration at two spacecraft. For self-similarly expanding ICMEs the expansion in size can be written as a power-law with an expansion factor $n_a$, observationally constrained to $0.45$\,$<$\,$n_a$\,$<$\,1.14 \citep[see e.g.][]{bothmer98,leitner07,demoulin08,gulisano12,2019ApJ...877...77V}. Accordingly, the falloff of the magnetic field magnitude is assumed to follow a power-law with an expansion factor $n_B$, observationally constrained to $0.88$\,$<$\,$n_a$\,$<$\,1.89.  The size of the MC was estimated assuming that the flux rope has a circular cross section and that the detector passes through the center of the flux rope. The ``size'' of the MC therefore represents the diameter of the flux rope. Assuming a Lundquist-type of flux rope, the axial and poloidal flux of the MC were calculated  using Equation~(52) in \cite{2000ApJ...539..954D} and Equation~(3) in \cite{2007ApJ...659..758Q}. The length of the flux rope at Earth was estimated according to \citet{2000ApJ...539..954D}, whereas the length at {\it Solar Orbiter} was estimated assuming the flux rope length expanded self-similarly according to the same power-law as the radial expansion. 

From Table~\ref{tab:in_situ} it can be seen that both power-law expansion factors, $n_a$ and $n_B$ for the size and magnetic field magnitude, respectively, reside at larger values of the observationally constraint interval, indicating a substantial expansion. The corresponding values of axial magnetic flux and helicity indicate that both magnetic flux and helicity are roughly conserved from {\it Solar Orbiter} to Earth, though uncertainties are rather large. Based on the assumption of self-similarity and using the estimated power-law index $n_B$, the average magnetic field magnitude of the flux rope at 10\,Mm above the solar surface can be estimated as $B_0$\,=\,200$\pm$300\,\gauss\ \citep[see, e.g., Equation (14) in][]{dumbovic18b}.  We note that the large uncertainty is due to error propagation of relatively large errors in the estimation of expansion factors $n_a$ and $n_B$, which are directly related to the uncertainty of estimation of the MC signature borders in the \textit{in-situ} measurements. 

\begin{table}[t]
\footnotesize
\caption{
\label{tab:in_situ}
{\it In-situ} measurements of the sheath, frontal region and MC at {\it Solar Orbiter} and Earth, as well as properties deduced for the ICME.}
\centering
\begin{tabular}{|l|c|c|}
\hline
	&	{\it Solar Orbiter}	&	OMNI	\\
\hline					
distance [AU]	&	$0.85 \pm 0.01$	&	$0.98 \pm 0.01$	\\
Shock arrrival [DOY]	&	$307.60 \pm 0.025$	&	$307.90 \pm 0.025$	\\
Frontal region start [DOY]	&	$308.00 \pm 0.025$	&	$308.30 \pm 0.025$	\\
Magnetic cloud start [DOY]	&	$308.30 \pm 0.025$	&	$308.55 \pm 0.025$	\\
Magnetic cloud end [DOY]	&	$308.85 \pm 0.025$	&	$309.15 \pm 0.025$	\\
\hline					
SHEATH	&		&		\\
\hline					
size [AU]	&	$0.15 \pm 0.01$	&	$0.16 \pm 0.01$	\\
$B_{\mathrm{avg}}$ [nT]	&	$25 \pm 1$	&	$17 \pm 1$	\\
$v_{\mathrm{avg}}$ [km/s]	&	$670 \pm 10$	&	$700 \pm 10$	\\
\hline					
FRONTAL REGION	&		&		\\
\hline					
size [AU]	&	$0.13 \pm 0.01$	&	$0.10 \pm 0.01$	\\
$B_{\mathrm{avg}}$ [nT]	&	$19 \pm 1$	&	$20 \pm 1$	\\
$v_{\mathrm{avg}}$ [km/s]	&	$730 \pm 10$	&	$710 \pm 10$	\\
\hline					
MAGNETIC CLOUD	&		&		\\
\hline					
$v_{\mathrm{lead}}$ [km/s]	&	$680 \pm 10$	&	$680 \pm 10$	\\
$v_{\mathrm{trail}}$ [km/s]	&	$570 \pm 10$	&	$570 \pm 10$	\\
size [AU]	&	$0.20 \pm 0.01$	&	$0.22 \pm 0.01$	\\
$B_{\mathrm{avg}}$ [nT]	&	$18 \pm 1$	&	$14 \pm 1$	\\
$v_{\mathrm{avg}}$ [km/s]	&	$625 \pm 7$	&	$625 \pm 7$	\\
$v_{\mathrm{exp}}$ [km/s]	&	$55 \pm 7$	&	$55 \pm 7$	\\
\hline					
ICME PROPERTIES	&		&		\\
\hline					
$n_a$	&	$1.1 \pm 0.6$	&	$1.1 \pm 0.6$	\\
$n_B$	&	$1.8 \pm 0.7$	&	$1.8 \pm 0.7$	\\
$B$ [$10^{-5}$ nT]	&	$1.8 \pm 0.1$	&	$1.4 \pm 0.1$	\\
$a$ [$10^{12}$ cm]	&	$1.5 \pm 0.1$	&	$1.6 \pm 0.1$	\\
flux rope length [$10^{12}$ cm]	&	$7 \pm 2$	&	$8 \pm 3$	\\
poloidal flux [$10^{21}$ Mx]	&	$0.75 \pm 0.26$	&	$0.74 \pm 0.25$ \\
axial flux [$10^{21}$ Mx]	&	$0.56 \pm 0.08$	&	$0.51 \pm 0.08$	\\
helicity [$10^{40}$ \mxmx]	&	$24 \pm 17$	&	$28 \pm 16$	\\
geometrical sense	&	left-handed	& left-handed \\
\hline
\end{tabular}
\normalsize
\end{table}

\section{Summary and Discussion}
\label{s:discussion}

In this paper we connect coronal observations and magnetic field-related properties of solar eruptivity to observations and modeling of the interplanetary consequences. A MC hit Earth on 2021~November~04 that was initiated by a sequence of solar eruptions hosted by AR~12891 on late November~01/early November~02: a C-class flare (SOL-2021-11-01T23:35C4.5) followed by a flare-less filament eruption and a double-peak M-class flare (SOL-2021-11-02T01:20M1.6 and SOL-2021-11-02T02:50M1.7). For an unambiguous ICME-MC association, we made use of the GCS reconstruction technique (based on simultaneous observations by SOHO/LASCO and STEREO-A coronagraphy, separated by $\sim$34~degrees with respect to the Sun-Earth line and at the time of arrival of the ICME at Earth) and by feeding the GCS-derived geometry (propagation direction and de-projected speed of the ICME) into DBEM. For the analysis of the MC we took particular advantage of multi-spacecraft observations of the almost perfectly radially aligned {\it Solar Orbiter} and ACE/Wind satellites (separated by only 3~degrees in longitude). This setting allowed us to trace magnetic-field related physical parameters (average field strength, magnetic flux and magnetic helicity) from the solar source region (approximated based on a time series of NLFF models) to a distance of 0.84~AU ({\it Solar Orbiter} MAG and SWA) and further to 0.98~AU (ACE/Wind), assuming that the ejected magnetic field structure did not change its magnetic topology during the transit through interplanetary space. The latter is supported by the fact that the corona and interplanetary space represent a frozen-in environment. Our main findings are summarized in Table~\ref{tab:fr_props} and discussed hereafter.

Our results suggest the coronal magnetic field which was subject to eruptive activity in the form of a sheared arcade/flux rope oriented along of the northern/southern portion of the flare-relevant PIL (Figure\,\href{fig:b_field}{\ref{fig:b_field}(a)} and \href{fig:b_field}{\ref{fig:b_field}(b)}) and supported by the spatial association to an observed filament and a sigmoid seen in EUV observations (Figure\,\href{fig:preflare_corona}{\ref{fig:preflare_corona}(b)} and \href{fig:preflare_corona}{\ref{fig:preflare_corona}(c)}). The NLFF modeling suggests that a left-handed (negative helicity) flux rope existed with the central axis/arcade field residing at coronal altitude of $\lesssim$5\,Mm/8--10\,Mm above photospheric levels (Figure\,\href{fig:b_field}{\ref{fig:b_field}(c)} and \href{fig:b_field}{\ref{fig:b_field}(e)}), the latter in close agreement with a stereoscopic estimate of the height of the observed filament (Figure\,\href{fig:preflare_corona}{\ref{fig:preflare_corona}(d)}). The stereoscopic estimates for different time instances moreover suggest that the filamentary material traced was  quickly replenished, allowing to conclude that the overall (supporting) magnetic field configuration of the filament must have on overall survived the repeated eruptive activity. The observed coronal EUV emission (Figure\,\href{fig:sdo_still}{\ref{fig:sdo_still}} and accompanying movie) as well as the time evolution of the coronal free energy and current-carrying helicity (Figure\,\href{fig:model_vs_obs}{\ref{fig:model_vs_obs}(c)} and \href{fig:model_vs_obs}{\ref{fig:model_vs_obs}(d)}, respectively) suggest that the different eruptive phenomena represent different episodes of a single yet spatially and temporally extended eruptive coronal process. In that course, as estimated from a time series of NLFF models, a total of $\approx$\,1.4$\pm$0.4$\times10^{31}$\,\erg\ in free magnetic energy and a total of $\approx$\,1.5$\pm$0.4$\times10^{41}$\,\mxmx\ in current-carrying helicity was released from the corresponding coronal budgets, supporting the scenario of a successful ejection of left-handed (negative-helicity) magnetic field.

An important question is that of whether the pre-existing coronal structure or the formation process by magnetic reconnection during the eruption process are more relevant in determining the structure reaching 1~AU. In the case of a relatively stronger importance of the eruptive process one would expect the magnetic flux of the MC to clearly exceed that of the (pre-) eruptive solar structure \citep[e.g.,][]{2004ApJ...602..422L}. This scenario is supported by our finding of an axial magnetic flux in the MC being approximately twice that of the solar source region (filament/flux rope), as is the case if we assume the cross section of the flux rope to span roughly 17\,Mm across the flare-relevant PIL (Figure\,\href{fig:b_field}{\ref{fig:b_field}(b)} -- \href{fig:b_field}{\ref{fig:b_field}(e)}). The realism of the assumed cross-sectional area is supported by the spatial distribution of intense axial electric current. This overall scenario --- of a considerable contribution of the eruptive process to the structural properties of the ejected flux rope --- is also supported by our finding of the helicity of the MC (approximated by a cylindrical LFF model)  to exceed that of the solar flux rope (approximated by the change to the current-carrying helicity due to the repeated eruptive activity). This is in line with the suggestion of \cite{2004JGRA..109.5106L} that helical field is added to that of the pre-eruptive structure in the course of the reconnection process. Contrasting their overall conclusion of the magnetic reconnection in the eruptive process being capable of explaining the MC helicity \citep[see also, e.g.,][]{2007SoPh..244...45L,2007ApJ...659..758Q}, our results --- given the rather large uncertainty ranges --- indicate that the helicity budget of the pre-eruptive structure may need to be considered as essential contribution too. 

\begin{table*}
\footnotesize
\caption{
\label{tab:fr_props}
Properties of the left-handed magnetic flux rope expelled from the solar source region (AR~12891) during an extended period of activity (a C4.5-class flare, a flare-less filament eruption, and a double-peak M-class flare) in comparison to that of the associated MC detected in near Earth.
}
\centering
\begin{tabular}{|l|c|c|}
\hline
~ & Solar Source Region (Low Corona) & Interplanetary CME/Magnetic Cloud ({\it Solar Orbiter} $\to$ Earth) \\
\hline
Distance from Sun (AU) & 0.00 & 0.85 $\to$ 0.98\\
Average magnetic field strength (\gauss) & & \\
\textit{-- Flux rope} & 262$\pm$126$^{(a)}$ (200$\pm$300$^{(b)}$) & (1.8$\pm$0.1)$\cdot10^{-4}$ $\to$ (1.4$\pm$0.1)$\cdot10^{-4}$	\\
Magnetic flux ($\times10^{21}$\,\mx) & & \\
\textit{-- Flux rope (axial)} & 0.3$\pm$0.2$^{(a)}$ & 0.6$\pm$0.1 $\to$ 0.5$\pm$0.1\\
\textit{-- Flux rope (poloidal)} & & 0.8$\pm$0.3 $\to$ 0.7$\pm$0.3\\
\textit{-- Flare ribbons (accumulated)} & 1.5$\pm$0.1 & \\ 
\textit{-- Coronal dimmings (accumulated)} & 9.7$\pm$0.5 & \\ 
Helicity (unsigned; $\times10^{41}$\mxmx) & & \\ 
\textit{-- Flux rope} & 1.5$\pm$0.4 & 2.4$\pm$1.7 $\to$ 2.8$\pm$1.6\\
\hline
\hline
\end{tabular}
\tablefoot{
$^{(a)}$Estimated from vertical planes as exemplary shown in Figure\,\href{fig:b_field}{\ref{fig:b_field}(c)}--\href{fig:b_field}{\ref{fig:b_field}(e)}, i.e., spanning $\sim$\,16\,$\times$\,17\,Mm$^2$. $^{(b)}$Back-extrapolation of {\it in-situ} estimate based on the assumption of self-similarity and using the estimated power-law index derived from in situ analysis of {\it Solar Orbiter} and OMNI measurements (for details see main text).
}
\normalsize
\end{table*}

From flare ribbon and coronal dimming emission we obtain the magnetic flux injected to the CME flux rope (Figure\,\href{fig:model_vs_obs}{\ref{fig:model_vs_obs}(d)} and \href{fig:model_vs_obs}{\ref{fig:model_vs_obs}(f)}, respectively). The reconnection processes started with the observed C4.5 flare, i.e., well before the flare-less filament eruption and M-class flaring, and spatially concentrated in the center of AR~12891. In comparison, the dimming-associated reconnection flux covering remote areas in the outskirts of the AR shows regions involved in the reconnection process --- first in the direction of the observed flare-less filament eruption (in the northern parts of the AR) and later, at the time when the CME fully erupted, towards solar south (Figure\,\href{fig:dtiming}{\ref{fig:dtiming}(a)} -- \href{fig:dtiming}{\ref{fig:dtiming}(d)}). Comparison of the relative timing of the cumulative reconnection fluxes reveals a period of significant contribution from dimmings without a simultaneous significant contribution from flare ribbons. This suggests that during the flare-less filament eruption and the early phase of the double-peak M-class flare, between 02~November $\sim$01:00--02:00\,UT, magnetic reconnection was taking place high in the solar corona, too weak to produce a measurable imprint onto the low solar atmosphere where the flare ribbons have been tracked (using AIA 1600\,\AA\ images). This is also supported by the fact that the flare ribbons were much more pronounced and exhibited larger spatial extents when observed in AIA 304\,\AA\ (compare left and middle columns in Figure\,\href{fig:sdo_still}{\ref{fig:sdo_still}}). Therefore, the ribbon-based cumulative estimate likely represents a lower limit of the total reconnection flux.

During the series of eruptive events that originated from AR~12891, given the estimated uncertainties, a total magnetic flux was accumulated in flare ribbons comparable to the poloidal flux estimated for the associated MC, in line with the findings in other case studies \citep[e.g.,][]{2004JGRA..109.5106L,2006SoPh..238..117A,2008AnGeo..26.3139M, 2017SoPh..292...93T} as well as the statistical study of \cite{2007ApJ...659..758Q}. In sharp contrast to those, however, is our finding of the dimming reconnection flux to exceed the poloidal flux of the associated MC by a factor of $\sim$10. The reconnection flux estimated from dimming regions facing the problem of projection effects due to the nature of EUV emission being LOS-integrated, and therefore representing only an upper limit for the actual flux processed through magnetic reconnection, may serve as a partial explanation for the large difference. Another partial explanation might be related to the fact that only the {\it in-situ} measured well-defined (``inner'') structure of the MC was used for considerations, while the (distorted) frontal region was disregarded (see Figure\,\ref{fig:insitu}). Upon inclusion of the frontal region in the calculations (i.e., considering a larger size and magnetic field magnitude since the magnetic field there is compressed), one would obtain a larger estimate of the magnetic fluxes. In that case, however, the assumption of a symmetric flux rope with a circular cross section would not be justified any longer, i.e., the Lundquist solution not applicable.

We do not have information on how the flux rope evolved between Sun and {\it Solar Orbiter}. However, back-extrapolation of specific magnetic-field related properties of the ICME/MC to the solar source region, under the assumption that the structure evolved in the same manner as it did between {\it Solar Orbiter} to Earth, allow a comparison to the remote solar observations and modeling. For instance, back-extrapolation of the average magnetic field strength of the MC at {\it Solar Orbiter} and Earth ((1.8$\pm$0.1)\,$\times$\,$10^{-4}$\,\gauss\ and (1.4$\pm$0.1)\,$\times$\,$10^{-4}$\,\gauss, respectively) suggest the corresponding solar source-region value as $\simeq$\,200$\pm$300\,\gauss. The overall agreement with the respective estimate from NLFF modeling ($\simeq$\,262$\pm$126\,\gauss) serves as a strong indication that the CME indeed evolved self-similarly between Sun and {\it Solar Orbiter}. In line, very well-matching {\it in-situ} signatures of the magnetic structure in {\it Solar Orbiter} and ACE/Wind with the corresponding spacecraft separated in longitude by less than 3~degrees, suggest an expansion of the ICME/MC structure with a power law index of 1.1$\pm$0.6. DBEM shows that {\it Solar Orbiter} and ACE/Wind probed the CME close to its apex (Figure\,\href{fig:icme}{\ref{fig:icme}(e)}). Hence, the interpretation of our results holds for the CME nose but might be different for flank regions. Nevertheless, the findings are in line with the known self-similar manner of expansion \citep[e.g.,][]{2019ApJ...877...77V} and the coherency in structure over longitudinal ranges of $\gtrsim$10~degrees \citep{2018ApJ...864L...7L}. 

\section{Conclusion}
\label{sec:conclusion}

Combining sophisticated numerical (NLFF) and geometrical reconstruction techniques (stereoscopy) with remote-sensing data at various distances from the Sun allows insights into the structural characteristics of the (pre-)eruptive solar source region and fluxes and helicities processed by means of magnetic reconnection (the latter estimated from flare ribbon and coronal dimming signatures). The spatial and propagation characteristics of the associated (I)CME can be reliably estimated from geometric fitting techniques (GCS) and interplanetary propagation models (DBEM). From the presented application it is evident that the {\it in-situ}  plasma and field measurements of the well-established ACE/Wind satellites in combination with that of the recently launched {\it Solar Orbiter} mission serves as an eligible combination to pursue in-depth studies of the ICME/MC characteristics between {\it Solar Orbiter} and Earth. However, the uncertainties especially in the derived magnetic-field-related parameters leave many questions open and prevent us from constraining the obtained results to unique interpretations, also due to the lack of {\it in-situ} data at closer distances to the Sun. The latter will be partially compensated by measurements of {\it Solar Orbiter} at later stages of the mission or with co-aligned measurements from Parker Solar Probe. 

\begin{acknowledgements}
We thank the referee for the positive evaluation of our manuscript and for the suggestions to optimize it.  JKT acknowledges the Austrian Science Fund (FWF): P31413-N27. MD acknowledges support by the Croatian Science Foundation under the project IP-2020-02-9893 (ICOHOSS). KD acknowledges support from NASA under award No. 80NSSC21K0738 and the NSF under AGS-ST Grant 2154653. MT gratefully acknowledges the Austrian Science Fund (FWF): P 33285. AMV and ED acknowledge the Austrian Science Fund (FWF): I4555. SDO data are courtesy of the NASA/SDO AIA and HMI science teams. We acknowledge and thank the ACE SWEPAM and MAG instrument teams and the ACE Science Center for providing the ACE data. {\it Solar Orbiter} is a mission of international cooperation between ESA and NASA, operated by ESA.
\end{acknowledgements}

\bibliographystyle{aa}
\bibliography{bibliography}
\end{document}